\documentclass[twocolumn, nac, superscriptaddress]{revtex4-1}
\usepackage{amssymb}
\usepackage{amsmath}
\usepackage{epsfig}
\usepackage{color}
\usepackage{graphics, graphicx}
\usepackage{bbold}
\usepackage{psfrag}
\usepackage{mathcomp}
\usepackage{subfigure}
\usepackage{verbatim}
\usepackage{color}
\usepackage[colorlinks,citecolor=blue]{hyperref}
\def\cp#1{\mathbf{#1}}

\begin{document}
\title{Engineering Non-Hermitian Skin Effect with Band Topology in Ultracold Gases}
\author{Lihong Zhou}
\affiliation{Beijing National Laboratory for Condensed Matter Physics, Institute of Physics, Chinese Academy of Sciences, Beijing 100190, China}
\author{Haowei Li}
\affiliation{CAS Key Laboratory of Quantum Information, University of Science and Technology of China, Hefei 230026, China}
\affiliation{CAS Center For Excellence in Quantum Information and Quantum Physics, Hefei 230026, China}
\author{Wei Yi}
\email{wyiz@ustc.edu.cn}
\affiliation{CAS Key Laboratory of Quantum Information, University of Science and Technology of China, Hefei 230026, China}
\affiliation{CAS Center For Excellence in Quantum Information and Quantum Physics, Hefei 230026, China}
\author{Xiaoling Cui}
\email{xlcui@iphy.ac.cn}
\affiliation{Beijing National Laboratory for Condensed Matter Physics, Institute of Physics, Chinese Academy of Sciences, Beijing 100190, China}
\affiliation{Songshan Lake Materials Laboratory , Dongguan, Guangdong 523808, China}
\date{\today}

\begin{abstract}
{\it \bf{Abstract}}

Non-Hermitian skin effect(NHSE) describes a unique non-Hermitian phenomenon that all eigen-modes are localized near the boundary, and has profound impact on a wide range of bulk properties. 
In particular, topological systems with NHSE have stimulated extensive research interests recently, given the fresh theoretical and experimental challenges therein. Here we propose a readily implementable scheme for achieving NHSE with band topology in ultracold gases. Specifically, the scheme realizes the one-dimensional optical Raman lattice with two types of spin-orbit coupling (SOC) 
and an additional laser-induced dissipation.
By tuning the dissipation and the SOC strengths, NHSE and  band topology can be individually controlled such that they can coexist in a considerable parameter regime.
To identify the topological phase in the presence of NHSE, we have restored the bulk-boundary correspondence by invoking the non-Bloch band theory, and discussed the dynamic signals for detection.
Our work serves as a guideline for engineering topological lattices with NHSE in the highly tunable environment of cold atoms, paving the way for future studies of exotic non-Hermitian physics in a genuine quantum many-body setting.
\end{abstract}

\maketitle

{\it \bf{Introduction}}

 Open quantum systems undergoing particle or energy loss can be effectively described by non-Hermitian Hamiltonians. They exhibit intriguing non-Hermitian phenomena that are absent in their Hermitian counterparts, and have thus attracted significant attention in recent years~\cite{PT_review,Uedareview}.
An outstanding example here is the non-Hermitian skin effect (NHSE)~\cite{Wang1, Wang2, Wang3, FoaTorres, Thomale, Murakami, Fang1, Sato, Fang2, Yang, Gong, Longhi, Longhi2, Chen,dengyi, Slager, Schindler1, Schindler2}, under which all bulk eigenstates are localized near the boundary. While NHSE is topologically protected by the winding of eigen-spectrum in the complex energy plane~\cite{Fang1,Sato}, the spectrum itself is sensitive to the actual boundary condition. For instance, both the eigen-spectrum and eigen-wavefunction can be dramatically different under an open boundary condition (OBC) from those under a periodic boundary condition (PBC).
A remarkable consequence is the failure of conventional bulk-boundary correspondence in topological systems with NHSE, whose restoration
calls for the so-called non-Bloch band theory by employing  a generalized Brillouin zone (GBZ)~\cite{Wang1,Wang2,Murakami, Fang2}. 
Apart from the fundamental impact on band topology,  NHSE can strongly influence many other bulk properties such as the dynamics~\cite{Wang3, Chen2, Longhi, Chen3, Zhang}, the parity-time symmetry~\cite{nonBlochPT1,nonBlochPT2} and localization~\cite{disorder1,disorder2}. 

To date, NHSE has been observed in various non-Hermitian one-dimensional (1D) topological systems including photonics~\cite{Xue,Szameit,Fan}, topoelectrical circuits~\cite{Thomale2}, and metamaterials~\cite{Coulais2}, wherein the non-Bloch bulk-boundary correspondence has also been confirmed~\cite{Xue, Thomale2, Szameit}.
In these studies, NHSE is predominantly achieved through non-reciprocal hopping, by simulating either the Hatano-Nelson model~\cite{HN} or the non-reciprocal Su-Schrieffer-Heeger model~\cite{Wang1}. At the moment, the study of NHSE deserves a substantial extension to a broader context. On one hand, the appearance of NHSE is not limited to these models. For instance, a simple spin rotation in momentum space can directly convert the non-reciprocal hopping to on-site dissipation~\cite{Lee, Wang1}, under which NHSE persists~\cite{FoaTorres, Wang3, Longhi, Yang, Gong}. On the other hand, given the existing experiments are either classical or on the level of single photons, it is desirable to engineer NHSE in a quantum many-body setting, which would offer exciting opportunities for investigating the interplay of NHSEs with many-body statistics and interactions.

Ultracold atomic gases, with highly controllable parameters, are an ideal candidate for the task. In this platform, through the photon-mediated Raman coupling technique, both the 1D and 2D spin-orbit couplings (SOC) have been realized~\cite{SOC_1d_1, SOC_1d_2, SOC_1d_3, SOC_1d_4, SOC_1d_5, SOC_2d_1, SOC_2d_2, SOC_2d_3, SOC_2d_4}, culminating in the successful generation of topological bands in optical  lattices~\cite{SOC_2d_3, SOC_2d_4, syntheticD_1, syntheticD_2, Jo1}.
Meanwhile, laser-induced atom loss has enabled the experimental realization of parity-time symmetry in ultracold atoms~\cite{Luo, Gadway,Jo2}, and a
very recent experiment manages to incorporate the SOC with laser-induced loss in a single setup~\cite{Jo2}.

\begin{figure*}[ht]
\centering
\includegraphics[width=13cm]{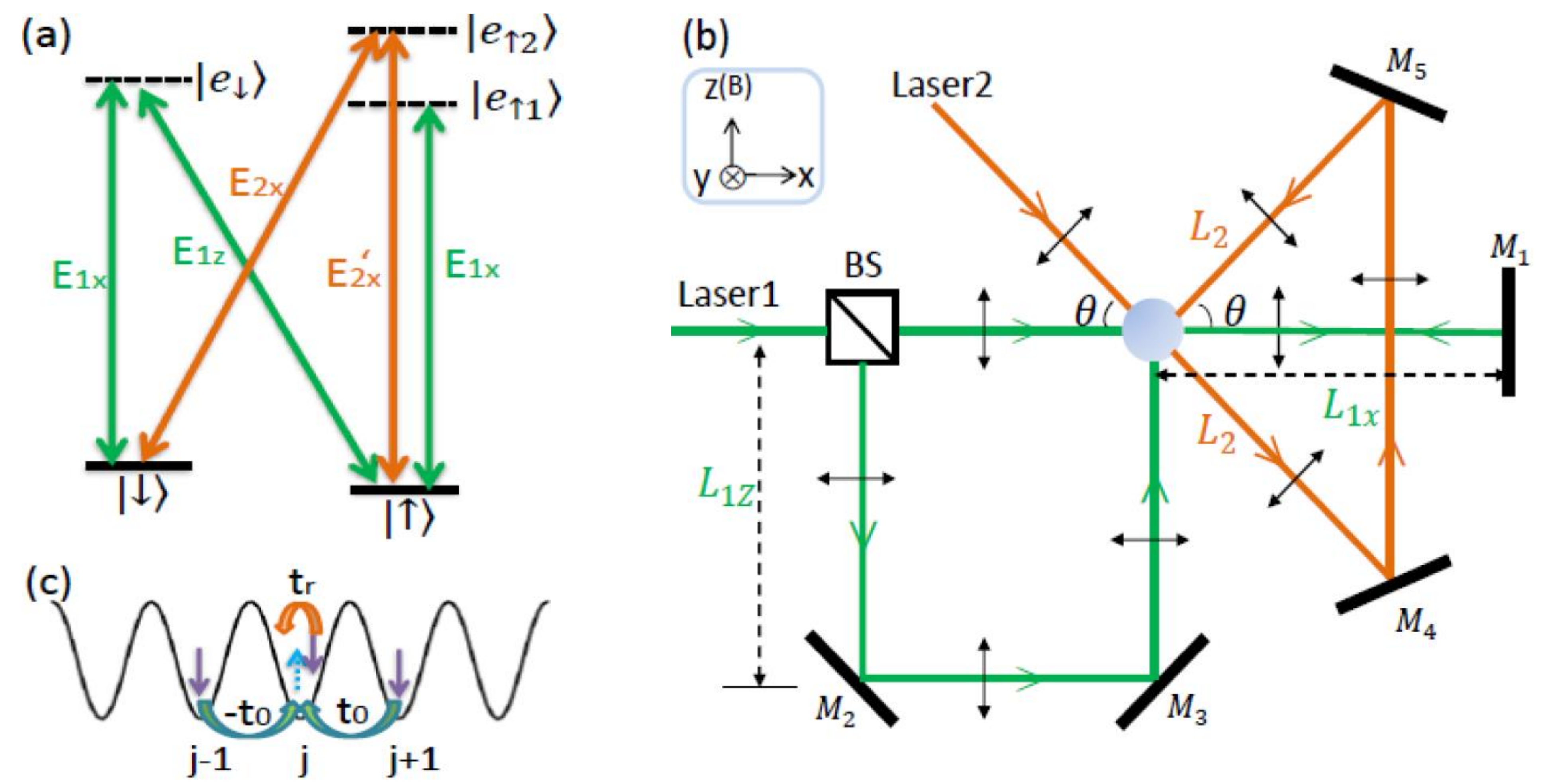}
\caption{{\bf{Schematics of the experimental setup in generating two types of spin-orbit couplings(SOCs)}}. (a) Two sets of Raman lasers couple the ground spin states via electronically excited states (dashed). (b) Laser configuration and optical-path diagram. The green and orange lines show the propagation directions of the two sets of lasers in (a), and the black arrows indicate their polarizations. 
 $L_{1x}, L_{1z}$ and $L_2$ are, respectively, the optical paths from the sample to mirror $M_1$, from the beam splitting to  mirror $M_2$, and from the sample to $M_4$ (and $M_5$). The optical lattice potential and the  $\Omega_0$-SOC are generated by  electric fields (${\cp E}_{1x},{\cp E}_{1z}$), and the $\Omega_r$-SOC is created by electric fields (${\cp E}_{2x},{\cp E}'_{2x}$).  (c) The $\Omega_0$- and $\Omega_r$-SOCs respectively generate the nearest-neighbor and the on-site spin flip with tunable strengths $\Omega_0e^{i\phi_0}(-1)^j$ and $\Omega_r e^{i\phi_rj}$ ($j$ is the site index).}  \label{fig_setup}
\end{figure*}

In light of these achievements, we propose to engineer NHSE with band topology in ultracold atoms by utilizing the Raman-assisted SOCs and laser-induced atom loss. Specifically, we consider a one-dimensional optical Raman lattice with two distinct types of SOCs,  where the non-trivial band topology is facilitated by one type of SOC, and the NHSE originates from the other, along with the laser-induced loss. While both types of SOC are indispensable in inducing the NHSE with non-trivial band topology,  they are not a trivial combination but exhibit strong interplay effect in this process.
We have mapped out the topological phase diagram, tabulated the parameter regimes for NHSE and band topology,
and proposed the dynamical detection scheme.
As all aspects of our proposal are readily accessible, our work represents a significant step toward the observation of NHSEs
and the associated exotic phenomena in a genuine quantum many-body setting.\\
{\it \bf{Results}}
\\
{\it \bf{Model}.}
As illustrated in Fig.~\ref{fig_setup}, we consider a two-component($\uparrow$,$\downarrow$) atomic gas in a 1D optical lattice (along $x$),
described by the Hamiltonian $H=\int dx H(x)$, with 
\begin{eqnarray}
H(x)&=&\frac{p_x^2}{2m}-V_0\cos(2k_0x) + M_0 \sin{k_0 x} (e^{i\phi_0}\sigma_++H.c.)\nonumber\\
&&+M_r(e^{i2k_rx}\sigma_++h.c.) + i\gamma\sigma_z. \label{hx}
\end{eqnarray}
Here $\sigma_{\pm}=\sigma_x\pm i\sigma_y$, with $\sigma_{\alpha}\ (\alpha=x,y,z)$ the Pauli matrices, and $\frac{p_x^2}{2m}$ is the kinetic term.

The optical Raman lattice~\cite{SOC_2d_3, SOC_2d_4, Jo1}, characterized by $V_0$ and $M_0$, is generated by two Raman lasers:
a standing wave propagating along $x$ with the electric-field vector ${\cp E}_{1x}={\cp e}_z 2E_{1x}e^{i\phi_{1+}} \cos(k_0x+\phi_{1-})$, where $\phi_{1\pm}=(\phi_{1x}\pm\phi'_{1x})/2$, and $\phi_{1x}\ (\phi'_{1x})$ is the phase of incident (reflected) light; and a plane wave propagating along $z$ with ${\cp E}_{1z}={\cp e}_x E_{1z} e^{ik_0z +\phi_{1z}}$.  As shown in Fig.~\ref{fig_setup}(b), $E_{1x}$ and $E_{1z}$ come from the same laser source (laser 1) through a beam splitter, enabling an easy manipulation of various relative phases.
For instance, the phase $\phi_{1-}(=-k_0L_{1x})$ is adjustable through the optical path $L_{1x}$ from the sample (grey dot) to mirror $M_1$. By taking $\phi_{1-}=-\pi/2$ and $z=0$, the field ${\cp E}_{1x}$ generates a lattice potential with spacing $a=\pi/k_0$ and lattice sites at $x_i=ia$. ${\cp E}_{1x},\  {\cp E}_{1z}$ combine to form the SOC that couples different spins with amplitude $M_0\sin(k_0x)$ and phase $\phi_0=\phi_{1+}-\phi_{1z}$. The phase $\phi_0=k_0(L_{1x}-2L_{1z})$ is tunable through $L_{1x}$
or $L_{1z}$, with $L_{1z}$ the perpendicular distance (along $\hat{z}$ between the mirrors $M_2,M_3$ and the sample.)

An additional SOC, characterized by $M_r$ and known as the equal Rashba and Dresslehaus coupling~\cite{SOC_1d_1, SOC_1d_2, SOC_1d_3, SOC_1d_4, SOC_1d_5, syntheticD_1, syntheticD_2}, is created by two plane-wave Raman lasers(${\cp E}_{2x}$,${\cp E}'_{2x}$) with opposite wave-vectors along $x$ and equal phase.  As shown in Fig.~\ref{fig_setup}(b), ${\cp E}_{2x}$ and ${\cp E}'_{2x}$ are both from the laser source 2 (with wave-vector $k_0$) and intersect at the sample following reflections by two mirrors $M_4,M_5$. By adjusting the angle $\theta$ between their propagation directions and $\hat{x}$, the recoil momentum  in (\ref{hx}) is tunable as $k_r=k_0\cos\theta$. Their relative phase can be tuned to zero, as is the case with (\ref{hx}), by adjusting the optical path $L$ between the mirrors and the sample such that $2k_0L_2(1+\sin\theta)=2n\pi$, with $n\in \mathbb{Z}$. 

 To ensure the atomic transitions as illustrated in Fig.~\ref{fig_setup}(a) and the zero detuning between different spins, it is required that the laser-frequency difference between $\{{\cp E}_{1x},\  {\cp E}_{1z}\}$, and that between $\{{\cp E}_{2x},\  {\cp E}'_{2x}\}$, exactly match the Zeeman splitting between $\uparrow$ and $\downarrow$. This can be achieved through the acousto-optical modulater (AOM), which has been widely used for two-photon Raman processes in cold atoms experiments~\cite{SOC_1d_1, SOC_1d_3, SOC_2d_1,SOC_2d_2,SOC_2d_3,Jo1}. Note that we have not shown AOM explicitly in Fig.~\ref{fig_setup}.

A laser-induced loss term, characterized by $\gamma$, is generated by coupling spin-$\downarrow$ atom to an excited state that is subsequently lost from the system due to spontaneous emission~\cite{Luo, Gadway,Jo2}. The conditional dynamics of the system under post-selection is characterized by the non-Hermitian Hamiltonian (\ref{hx}), after dropping the global loss term ($\sim i\gamma$). 

\begin{figure*}[ht]
\centering
\includegraphics[width=12cm]{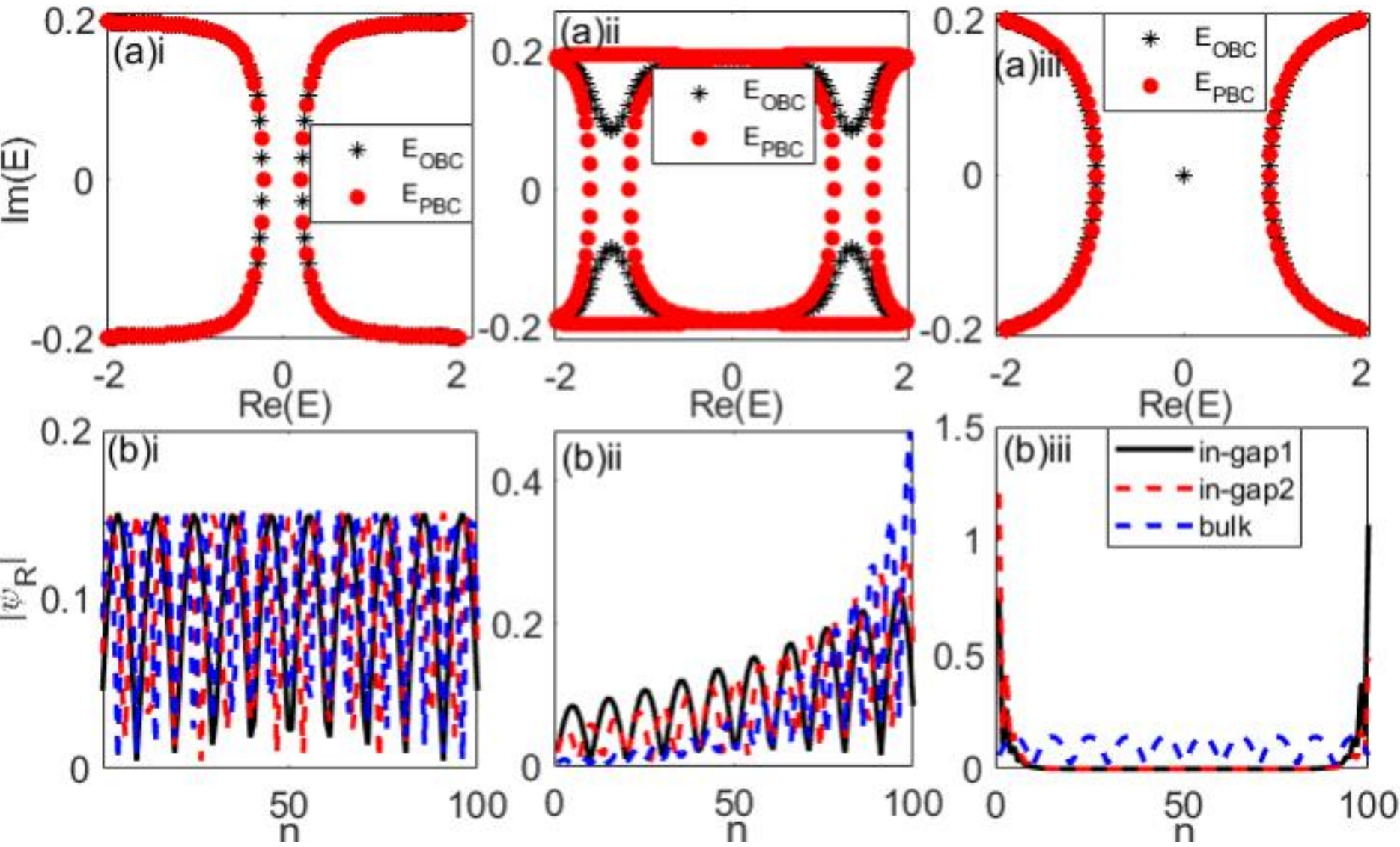}
\caption{ {\bf{Eigenspectrum ((a)i-iii) and spatial profile of eigen-modes ((b)i-iii) for case-I and -II when only one spin-orbit coupling (SOC) is present}}. Here we take the dissipation strength $\gamma=0.2$, and the other parameters are $(\Omega_0,\phi_0,\Omega_r,\phi_r)=(0,-,0.3,\pi)$ for ((a)i,(b)i); $(0,-,0.3,\pi/2)$ for ((a)ii,(b)ii); $(0.5,\pi/2,0,-)$ for ((a)iii,(b)iii).
The non-Hermitian skin effect shows up only in ((a)ii,(b)ii).
((a)iii,(b)iii) are topological with two in-gap zero-modes localized at both boundaries. Here the energy unit is taken as hopping $t$.} \label{fig_caseI_II}
\end{figure*}

The tight-binding model corresponding to Eq.(\ref{hx}) is
\begin{align}
H&=-t\sum_j(c_{j\uparrow}^\dagger c_{j+1\uparrow}+c_{j\downarrow}^\dagger c_{j+1\downarrow}+h.c.)\nonumber\\
&+\Omega_0\sum_j[e^{i\phi_0}(-1)^j(c_{j\uparrow}^\dagger c_{j+1\downarrow}-c_{j\uparrow}^\dagger c_{j-1\downarrow})+h.c.]\nonumber\\
&+\Omega_r\sum_j(e^{i\phi_rj}c_{j\uparrow}^\dagger c_{j\downarrow}+h.c.) +i\gamma\sum_j(c_{j\uparrow}^\dagger c_{j\uparrow}-c_{j\downarrow}^\dagger c_{j\downarrow}). \label{H}
\end{align}
Here the two SOCs respectively provide the nearest-neighbor and the on-site spin flip with amplitudes $\Omega_0$ and $\Omega_r$. We henceforth denote them as the $\Omega_0$- and $\Omega_r$-SOC, respectively,
along with the phase parameters $\phi_0(=\phi_{1+}-\phi_{1z})$ and $\phi_r(=2\pi k_r/k_0)$. 
Since all the parameters in (\ref{H}) are highly tunable, in this work we fix the hopping rate $t$ as the energy unit $(t=1)$ and take $\phi_0,\phi_r\in[0,2\pi)$.

Our scheme is applicable to a wide range of alkali and alkali-earth(-like) atoms. A promising candidate is $^{173}$Yb, where both the optical Raman lattice with band topology~\cite{Jo1} and non-Hermitian SOC~\cite{Jo2} have been realized using the Raman-induced $^{1}S_0\ \leftrightarrow \ ^{3}P_1$ transitions.  
A detailed Raman-transition scheme corresponding to Fig.~\ref{fig_setup}(a) for $^{173}$Yb can be found in [Supplementary Note 1].

\noindent
{\it \bf{NHSE and band topology}.} To provide insight and highlight the individual role of two SOCs, we consider the following three cases: 

\noindent
{\bf Case-I: $\Omega_0=0$, $\Omega_r\neq 0$}

This is the lattice version of the continuum gas with non-Hermitian SOC, as implemented recently in Ref.~\cite{Jo2}. Here, following the gauge transformation 
$c_{j\downarrow}\rightarrow c_{j\downarrow}e^{-i\phi_r j}$, we obtain the Bloch Hamiltonian $H(k)= -2t \cos(\phi_r/2) \cos \tilde{k}+\Omega_r\sigma_x+(i\gamma-2t\sin(\phi_r/2)\sin \tilde{k})\sigma_z$ (with $\tilde{k}=k+\phi_r/2$) and the eigenenergy
\begin{align}
E^{(I)}_{k\pm}&= -2t \cos(\phi_r/2) \cos \tilde{k} \nonumber\\
&\pm \sqrt{\Omega_r^2+(i\gamma-2t\sin(\phi_r/2)\sin \tilde{k})^2}. \label{E1}
\end{align}
The eigenspectrum supports two exceptional points at $\Omega_r=\gamma$, when $\tilde{k}=0,\pi$ or $\phi_r=0$.

Importantly, the system hosts NHSE for $\phi_r\neq 0,\pi$, as clearly indicated by the closed-loop topology~\cite{Fang1,Sato} of (\ref{E1}) in the complex plane. The presence of NHSE leads to distinct spectra under PBC and OBC, and localized bulk eigen-modes near the OBC boundary (see Fig.~\ref{fig_caseI_II} {((a)ii,(b)ii))}.
Nevertheless, this case is trivial in band topology, since $H(k)$ does not show any spin-winding as $k$ traverses the Brillouin zone. As a result, the OBC spectrum does not feature any in-gap topological modes (see Fig.~\ref{fig_caseI_II} {((a)i,(a)ii))}.

\noindent
{\bf Case-II: $\Omega_0\neq 0$, $\Omega_r=0$}

This is the case with only optical Raman lattice and dissipation. Following the transformation $c_{j\downarrow}\rightarrow (-1)^jc_{j\downarrow}$, the Bloch Hamiltonian $H(k)= 2\Omega_0\sin k (\sin\phi_0 \sigma_x +\cos\phi_0 \sigma_y)+(i\gamma-2t\cos k)\sigma_z$, and the eigenspectrum is
\begin{equation}
E^{(II)}_{k\pm}= \pm \sqrt{(2\Omega_0\sin k)^2+(i\gamma-2t\cos k)^2}. \label{E2}
\end{equation}
Clearly, there is no NHSE --- the spectrum (\ref{E2}) exhibits no loop structures in the complex plane. However, $H(k)$ possesses non-trivial band topology, as further confirmed in Fig.~\ref{fig_caseI_II} {((a)iii,(b)iii)} by the appearance of in-gap zero modes under OBC and their localized wave functions near boundaries.
The topological transition occurs at $\gamma=2\Omega_0$ for all $\phi_0$, when the band gap closes at $k=\pi/2,\ 3\pi/2$. We note that the topological phase of a similar model under a real Zeeman field  ($i\gamma\rightarrow \Gamma_z$) and $\phi_0=0$ has been studied in Ref.~\cite{Liu2014}, where the topological transition occurs at $\Gamma_z=2t$.

One can see that the $\Omega_r$-SOC and the $\Omega_0$-SOC can respectively give rise to  NHSE and band topology, as they respectively lead to spectral and wave-function windings. In order to achieve both in a single setting, one needs to incorporate all essential gradients to satisfy both winding conditions. A natural contender is by combining both types of SOCs, as well as the on-site loss.

\noindent
{\bf Case-III: $\Omega_0\neq 0$, $\Omega_r\neq0$}

\begin{figure}[h]
\includegraphics[width=8cm, height=4cm]{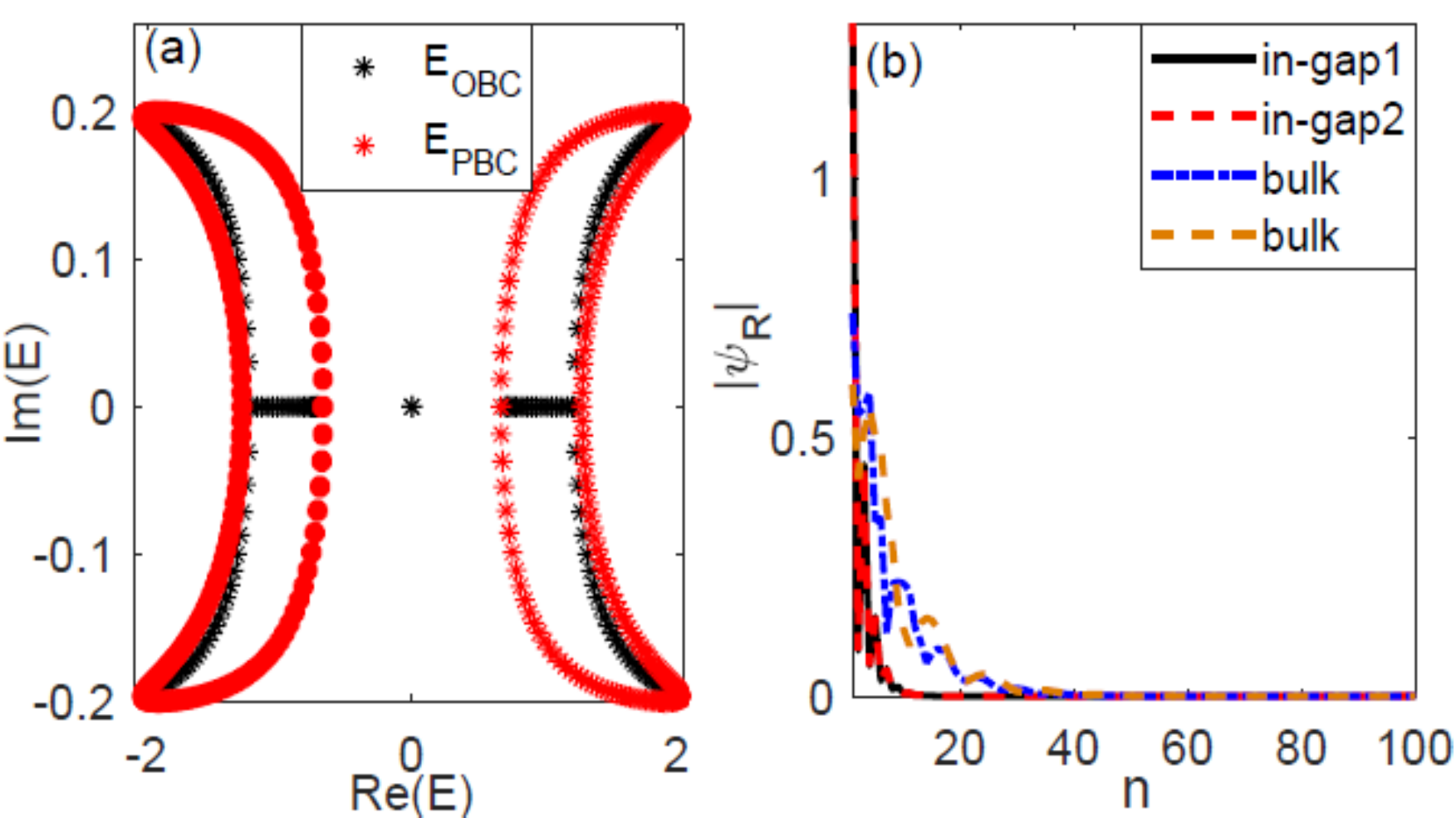}
\caption{{\bf{Coexistence of non-Hermitian skin effect and band topology for case-III with $\phi_r=\pi, \phi_0=\pi/2$, and $\Omega_0=0.5, \Omega_r=0.3, \gamma=0.2$}}. (a)Energy spectra under periodic boundary condition and open
boundary condition in the complex plane. (b) Spatial wave functions for the two topological edge modes and two randomly chosen bulk states. Here the energy unit is taken as hopping $t$.}
\label{fig_caseIII}
\end{figure}

When both types of SOCs are switched on, an analytical form of the eigenspectrum is generally unavailable. An exception is when $\phi_r=\pi$, under which the two SOCs are commensurate and $k$ is still a good quantum number. The Bloch Hamiltonian is
$H(k)= (\Omega_r+2\Omega_0 \sin\phi_0 \sin k) \sigma_x +2\Omega_0\cos\phi_0\sin k \sigma_y+(i\gamma-2t\cos k)\sigma_z$, with the eigenspectrum
\begin{align}
E^{\rm (III,\phi_r=\pi)}_{k\pm}&= \pm \left[(\Omega_r^2+(2\Omega_0\sin k)^2+4\Omega_r\Omega_0\sin\phi_0 \sin k \right.\nonumber\\
&\left.\ \ \ \ +(i\gamma-2t\cos k)^2\right]^{1/2}. \label{E3}
\end{align}
We see immediately that once $\phi_0\neq 0,\pi$, the spectrum (\ref{E3}) would form a loop in the complex plane, signifying the presence of NHSE. Moreover, $H(k)$ in this case keeps a similar spin winding pattern as in case-II, and thus the system acquires a band topology. In particular, when $\phi_0=\pi/2, \ 3\pi/2$ the system has the chiral symmetry: $\sigma_y H(k) \sigma_y =-H(k)$, which protects the degenerate topological zero modes. In Fig.~\ref{fig_caseIII}, we have numerically verified the coexistent skin and topological properties for $\phi_r=\pi$ and $\phi_0=\pi/2$, from both the different eigen-spectra between PBC and OBC
(Fig.~\ref{fig_caseIII}(a)), and the localized bulk state and in-gap zero modes
(Fig.~\ref{fig_caseIII}(b)).

\begin{figure}[t]
\includegraphics[width=8cm]{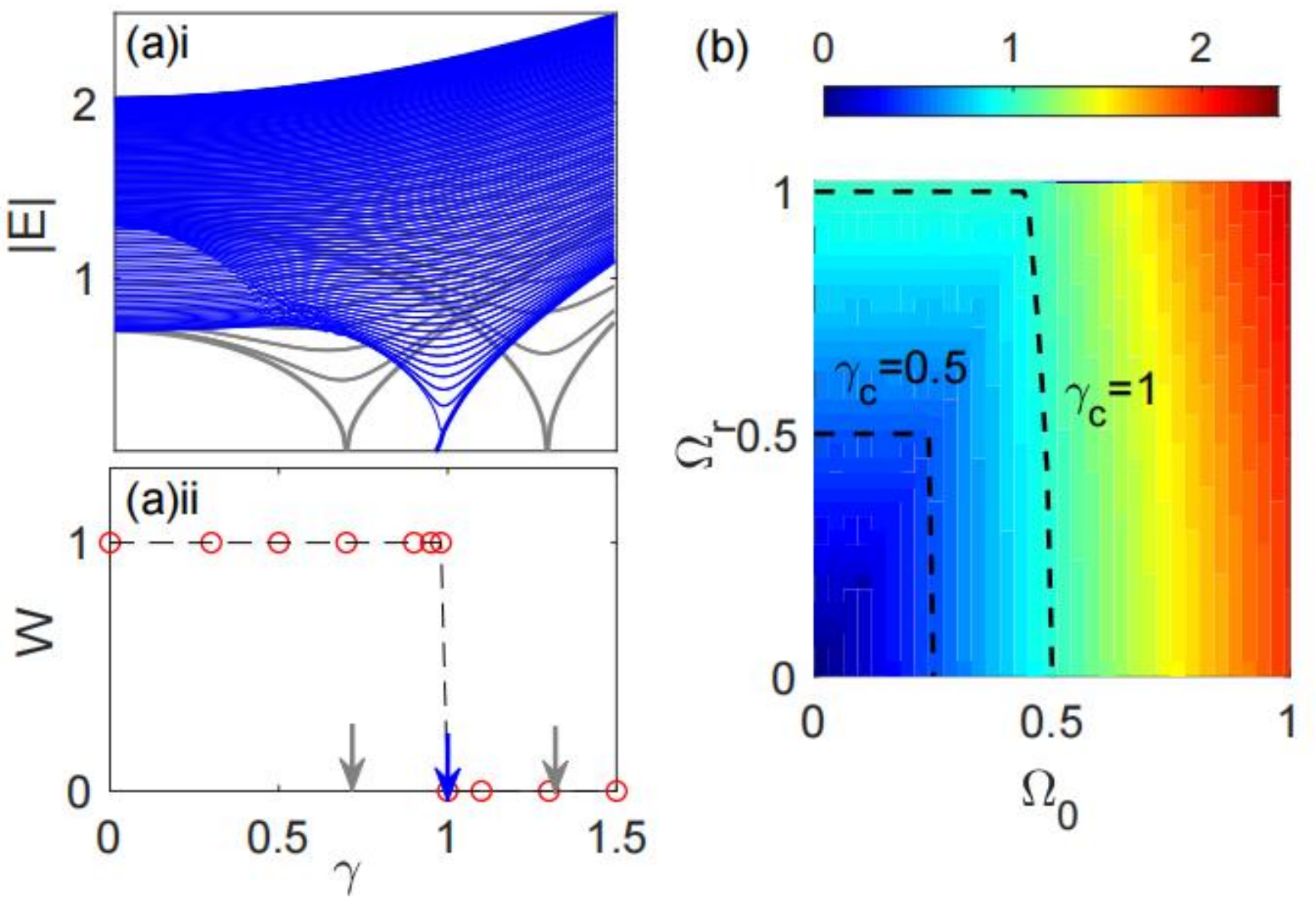}
\caption{{\bf{ Topological phase transition for case-III with $\phi_r=\pi,\ \phi_0=\pi/2$}}. { ((a)i,(a)ii)} The amplitude of spectrum $|E|$ under open
boundary condition  (blue) and the winding number $W$ obtained from ed Brillouin zone as functions of $\gamma$. Here we take $\Omega_0=0.5, \ \Omega_r=0.3$. The topological transition occurs at $\gamma_c\sim 1$, which differs from the periodic boundary condition predictions $\gamma_{c,PBC}=0.7,1.3$ (gray arrows). The periodic boundary condition eigenspectrum is shown in grey for comparison. { (b)} Contour plot of $\gamma_c$ in $(\Omega_0, \ \Omega_r)$ plane  (here the color bar represents the value of $\gamma_c$). The energy unit is taken as hopping $t$.} \label{fig_caseIII_topo}
\end{figure}

The band topology can be conveniently tuned by $\gamma$ and $\Omega_0,\Omega_r$. In Fig.~\ref{fig_caseIII_topo}{((a)i)}, we take a specific set of $\Omega_0$ and $\Omega_r$, and show that by increasing $\gamma$ to a critical $\gamma_c$, the in-gap zero modes merge into the bulk and the gap closes and reopens across this critical point. This signifies a topological transition into the trivial phase. Remarkably, $\gamma_c$ is different from the gap-closing point of the eigen-spectrum (\ref{E3}) under PBC, where $\gamma_{\rm c,PBC}=|\Omega_r\pm 2\Omega_0|$. This is exactly the breakdown of conventional bulk-boundary correspondence under NHSE.

To restore the bulk-boundary correspondence, we adopt the non-Bloch band theory~\cite{Wang1,Wang2,Murakami, Fang2} and compute the winding number $W$ in the GBZ. As shown in Fig.\ref{fig_caseIII_topo}{((a)ii)}, $W$ can well predict the topological transition under OBC: the in-gap zero modes emerge where $W=1$ ($\gamma<\gamma_c$) and vanish where $W=0$ ($\gamma>\gamma_c$). 
In fact, $\gamma_c$ can be obtained analytically through the gap-closing condition of the GBZ spectrum[Methods]. 
The resulting topological phase diagram is given in Fig.~\ref{fig_caseIII_topo} {(b)}, where $\gamma_c$ is plotted as functions of $\Omega_0$ and $\Omega_r$ for the case of $\phi_0=\pi/2$. The $\phi_0=3\pi/2$ case is found to share the same diagram due to symmetries[Methods].

When $(\phi_r, \phi_0)$ deviate from $(\pi,\pi/2)$ and $(\pi, 3\pi/2)$, the chiral symmetry is broken and the  topological modes would split and gradually merge into the bulk. In comparison, NHSE is much more robust, which can persist for all $(\phi_r,\phi_0)$ except for $\phi_r=0$ and two discrete points $(\pi,\pi)$ and $(\pi,0)$[Supplementary Note 2]. In Table~\ref{table}, we summarize the conditions for band topology and NHSE for cases I-III. A main message is that both SOCs are indispensable in achieving NHSE and topology simultaneously.   Their combination shows an intriguing interplay effect, instead of being a trivial superposition. For instance, the application of $\Omega_r$-SOC in case III changes the topology condition as compared to case II, and the   $\Omega_0$-SOC changes the skin condition as compared to case I.  In particular, for the latter case, the presence of $\Omega_0$-SOC further broadens the parameter region of observing skin modes, signifying a positive effect of band topology in enhancing NHSE.



\begin{widetext}

\begin{table}[h]
\begin{tabular}{|c|c|c|c|c|}
  \hline\hline
    & & topological & NHSE & topological + NHSE\\
  \hline
  case-I & $\Omega_0=0,\ \Omega_r\neq 0$ & $\times$ & $\phi_r\neq 0,\pi$ & $\times$ \\
  \hline
  case-II  & $\Omega_0\neq0,\ \Omega_r= 0$ & $\gamma<2\Omega_0$ & $\times$ & $\times$\\
  \hline
  case-III & $\Omega_0\neq0,\ \Omega_r\neq 0$ & $\phi_r=\pi, \phi_0=\frac{\pi}{2},\frac{3\pi}{2}$ and $\gamma<\gamma_c$ &  $\phi_r\neq 0$ and $(\phi_r,\phi_0)\neq (\pi,0), (\pi,\pi)$ & $\phi_r=\pi, \phi_0=\frac{\pi}{2},\frac{3\pi}{2}$ and $\gamma<\gamma_c$ \\
  \hline\hline
\end{tabular}
\caption{{\bf{Conditions for achieving topological phase and non-Hermitian skin effect (NHSE)} for various cases}.   $\Omega_0$ and $\Omega_r$ are respectively the nearest-neighbor and the on-site spin flip amplitudes, $\gamma$ is the dissipation strength, and $\phi_r, \phi_0$ are the phase parameters. "$\times$" means absence for all occasions. } \label{table}
\end{table}

\end{widetext}

\begin{figure}[h]
\includegraphics[width=8cm]{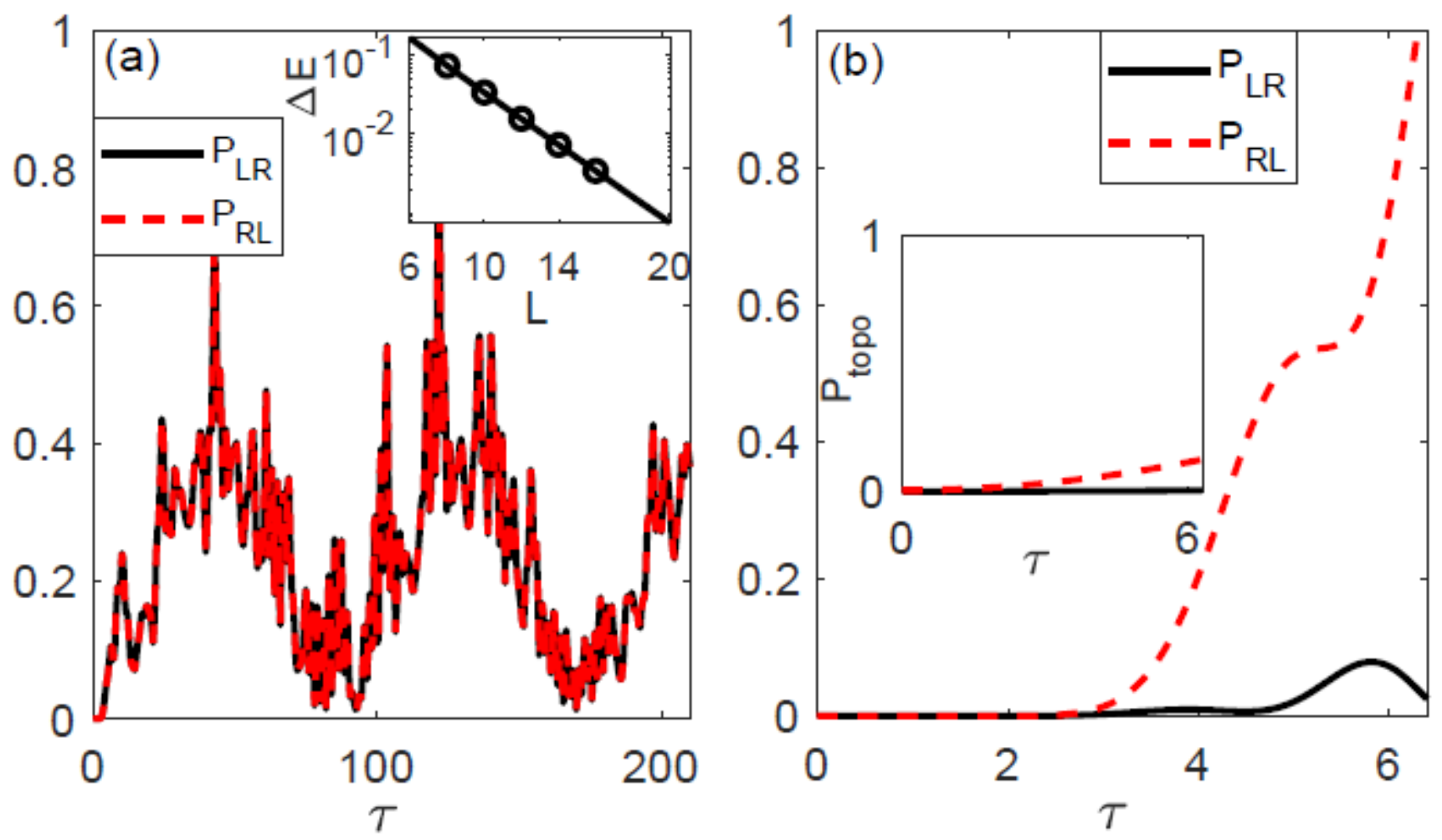}
\caption{{\bf{Edge-to-edge transport properties for the topological system without (a) and with (b) non-Hermitian skin effect}}. We take $\Omega_0=0.5,\ \Omega_r=0.3$, and $\gamma=0$ (a), $0.2$ (b).  In the main plots, the system size $L=8$.
The inset of (b) shows the contribution from the topological edge states. Here the energy unit is hopping $t$, and the time unit is $\hbar/t$.  }\label{fig_transport}
\end{figure}
\noindent
{\it \bf{Dynamic detection}.} To detect the topological phase with NHSE, we propose an edge-to-edge transport measurement. In the Hermitian case, the topological edge modes play the dominant role in such transport
~\cite{transport_expt}. However,  in the presence of NHSE, the transport is expected to be significantly modified since all bulk modes also localize near the edge. To examine such an effect, we compare two topological systems in our setup, one is Hermitian at $\gamma=0$, and the other is with NHSE at finite $\gamma$. We study the probability of particle occupation at the $\beta$-edge of the system at time $\tau$, when the initial state starts from the $\alpha$-edge ($\alpha,\beta=$L or R)
\begin{equation}
P_{\alpha\beta}(\tau) = |\langle \beta | e^{-iH\tau} |\alpha\rangle|^2.
\end{equation}

To eliminate the difference caused by spin, we take the initial state as the equal population of $\uparrow$ and $\downarrow$ and show its dynamics in Fig.~\ref{fig_transport}. In the Hermitian case (Fig.~\ref{fig_transport} (a)), the left-to-right ($P_{LR}$) and right-to-left ($P_{RL}$) transports are identical. As a manifestation of the topological edge states, the oscillation frequency of $P_{LR}$ (or $P_{RL}$), as given by the energy gap between the two edge modes in a finite-size system, is found to decay exponentially with increasing system size (inset of Fig.~\ref{fig_transport} (a)).
In the presence of NHSE (Fig.~\ref{fig_transport} (b)), the transport properties are dramatically different. Due to the localization of skin modes at the left boundary, the transport shows strong directional preference towards the left side, namely.
In this case, the topological edge modes play little role in affecting the dynamics (inset of Fig.~\ref{fig_transport} (b)). These features distinguish the topological phases with and without NHSE.

 Optical lattices with sharp boundaries can be implemented using box-trap potentials~\cite{Hadzibabic, Dalibard, Zwierlein, Moritz, Schmiedmayer}, where the spatial extent of an edge is determined by the optical wavelength $\sim 1\mu$m (much smaller than typical trap length $\sim 10$-$100\mu $m). Such a small imperfection does not visibly change the dynamics in Fig.~\ref{fig_transport}[Supplementary Note 3].  Even sharper edges ($\sim 10$nm) can be created via the dark state in atomic $\Lambda$-system\cite{Zoller2016, Gorshkov2016, nanoedge_expt}.
 Alternatively, without edges the NHSE can also manifest itself in bulk dynamics~\cite{Longhi, Chen2, Zhang}.  Indeed we have  confirmed that  NHSE can lead to visible directional bulk transport under typical harmonic trapping potentials[Supplementary Note 3].

For the detection of the non-Bloch band topology, one may further resort to quench dynamics or measurement of the biorthogonal chiral displacement~\cite{nbquench,qwnbquench}. Alternatively, topological edge states may be probed through a time-integrated state tomography~\cite{Xue}.\\

\par
\noindent
{\it \bf{Discussion}}

\noindent
We have proposed a realistic scheme in utilizing the SOCs and the spin-dependent dissipation in ultracold atoms to engineer NHSE with band topology.
We emphasize that both SOCs are indispensable in the scheme. Their mutual interference, along with their interplay with the on-site dissipation, determine the ultimate parameter regime for skin modes with non-trivial band topology (Table~\ref{table}). 

For future studies, an intriguing possibility would be tuning $\phi_r$ away from $\pi$ such that the two SOCs become incommensurate. The competition between quasiperiodicity and NHSE would potentially lead to unique localization features~\cite{disorder1,disorder2}.
Further generalization of our scheme to higher dimensions would offer the opportunity for
achieving Weyl exceptional rings~\cite{Xu}, and high-order skin effect and band topology featuring corner or hinge modes~\cite{hoti,hoskin,critskin}.
Moreover, the implementation of NHSE in ultracold atoms paves the way for exploring collective phenomena therein due to inter-atomic interactions, which are easily tunable through Feshbach resonances.
Our proposal therefore ushers in a wide variety of possibilities for the quantum simulation of non-Hermitian physics.\\

\noindent

{\it \bf{Methods}}
\\
{{\it \bf{Derivation of  tight-binding model:}}
To derive the tight-binding model, we expand the field operator $\psi_{\sigma}(x)=\sum_{i}\omega_{n=0}(x-x_i)c_{i\sigma}$, where $\omega_{n=0}(x)$ is the lowest-band Wannier function and $i$ is the index of lattice sites. In this way, the second-quantized single-particle Hamiltonian can be reduced to the tight-banding model, with the following parameters:

(1) the nearest-neighbor hopping term
\begin{align}
t&
=-\int dx  \omega_0(x)(\frac{p_x^2}{2m}-V_0\cos(2k_0x))\omega_0(x-a).
\end{align}

(2) the on-site spin-flip terms
\begin{align}
 &t_{\uparrow\downarrow}^{j}=\int dx\omega_{0}^\ast(x-x_j) M_r e^{i2k_r x}\omega_{0}(x-x_j)\equiv e^{i\phi_r j}\Omega_r,\\
 &t_{\downarrow\uparrow}^{j}=\int dx\omega_{0}^\ast(x-x_j) M_r e^{-i2k_r x}\omega_{0}(x-x_j)\equiv e^{-i\phi_r j}\Omega_r,
 \end{align}
where the amplitude $\Omega_r=M_r \int dx \omega_{0}^2(x) e^{i2k_r x}$, and the corresponding phase  $\phi_r=2k_r a=2\pi k_r/k_0$. Note that here we pin down the coordinate of the j-site atom as $x_j=ja$, with $a=\pi/k_0$ the lattice spacing.

(3) the nearest-neighbor spin-flip terms
\begin{align}
 &t_{\uparrow\downarrow}^{j,j+1}=\int dx\omega_{0}^\ast(x-x_j)(M_0 \sin{k_0 x}e^{i\phi_0})\omega_{0}(x-x_{j+1})\nonumber\\
 &\equiv e^{i\phi_0}(-1)^{j}\Omega_0,\\
 &t_{\uparrow\downarrow}^{j,j-1}=\int dx\omega_{0}^\ast(x-x_j)(M_0 \sin{k_0 x}e^{i\phi_0})\omega_{0}(x-x_{j-1})\nonumber\\
& \equiv -e^{i\phi_0}(-1)^{j}\Omega_0,
 \end{align}
where the amplitude $\Omega_0=M_0\int dx \omega_{0}(x)\sin{k_0 x}\omega_{0}(x-a)$. Again we have used $x_j=ja$.

Finally we get the tight-binding model as Eq.~(2) in the main text.
\\
{\it \bf{Non-Bloch band theory.}}
We have adopted the non-Bloch band theory~\cite{Wang1,Wang2,Murakami, Fang2} to investigate the  topological properties for case-III with non-Hermitian skin effect. Replacing the vector $k$  by $\beta=e^{ik}$ in the Bloch Hamiltonian,  the non-Bloch Hamiltonian in the spin space can be written as
 \begin{equation}
 H(\beta)=\left(
  \begin{array}{cc}
   i\gamma-t(\beta+\beta^{-1})&\Omega-e^{i\phi_0}t_{so}(\beta-\beta^{-1})\\
   \Omega+e^{-i\phi_0}t_{so}(\beta-\beta^{-1})& -i\gamma+t(\beta+\beta^{-1})
  \end{array}
\right).
\end{equation}

Then the eigenvalues $E$ follow
\begin{align}
E^2=&[i\gamma-t(\beta+\beta^{-1})]^2+[\Omega-e^{i\phi_0}t_{so}(\beta-\beta^{-1})]\nonumber\\
&(\Omega+e^{-i\phi_0}t_{so}(\beta-\beta^{-1})). \label{E_beta}
\end{align}
For a given value of $E$, the four solutions of $\beta$ can be organized as $|\beta_1|\le |\beta_2|\le|\beta_3|\le|\beta_4|$. Imposing the condition $|\beta_2|=|\beta_3|$ would pin down all$\beta$-solutions for the generalized Brillouin zone (GBZ).

The non-Bloch  winding number accumulated in the GBZ is then
\begin{equation}
W=\frac{i}{2\pi} \int_{\beta} \sum_{\nu=\pm} \langle u_{\nu L}(\beta) | \partial_\beta | u_{\nu R}(\beta)\rangle,
\end{equation}
where the right and left eigenvectors are defined through $H(\beta)|u_{\nu R}\rangle=E_{\beta\nu} |u_{\nu R}\rangle$ and $H^{\dag}(\beta)|u_{\nu L}\rangle=E_{\beta\nu}^* |u_{\nu L}\rangle$. Note that $H(\beta)$ satisfies the chiral symmetry $\sigma_y H(\beta)\sigma_y=-H(\beta)$, for $\phi_0=\pi/2,3\pi/2$.
\\
{\it \bf{Topological phase transition point.}}
The gap-closing condition, indicative of the topological transition, requires  the solution $E=0$ of (\ref{E_beta}), i.e.,
\begin{align}
0=&[i\gamma_c-t(\beta+\beta^{-1})]^2+[\Omega-e^{i\phi_0}t_{so}(\beta-\beta^{-1})]\nonumber\\
&[\Omega+e^{-i\phi_0}t_{so}(\beta-\beta^{-1})].
\end{align}
In combination with the continuum band requirement: $|\beta_1|\leq|\beta_2|=|\beta_3|\leq|\beta_4|$, we obtain the relation between $\gamma_c$ and $\Omega_0,\Omega_r$. In particular, since the topological zero modes are protected by the chiral symmetry when $\phi_0=\pi/2, \ 3\pi/2$, in the following we shall discuss the topological transition in these two cases separately.

{\bf{1. $\phi_0=\pi/2$}}

The solution of $\gamma_c$ as a function of $\Omega_0,\Omega_r$ can be divided to two regimes
\begin{align}
& {\rm (1)\quad for} \ \Omega_{0}\le\Omega_{0c}: \gamma_c=\Omega_r;\\
\label{ex1}
& {\rm (2)\quad for} \ \Omega_{0}\ge\Omega_{0c} : 2(t\gamma_c+\Omega_r\Omega_0)=(t+\Omega_0)
\nonumber\\
&\sqrt{(\Omega_r+\gamma_c)^2+4t^2-4\Omega_0^2}
-(t-\Omega_0)
\sqrt{(\Omega_r-\gamma_c)^2+4t^2-4\Omega_0^2}.
\end{align}
Here $\Omega_{0c}$ satisfies
\begin{equation}
(t+\Omega_{0c})(\sqrt{\Omega_r^2+t^2-\Omega_{0c}^2}-\Omega_r)=(t-\Omega_{0c})\sqrt{t^2-\Omega_{0c}^2}.
\end{equation}

{\bf{2. $\phi_0=3\pi/2$}}

From Eq.~(\ref{E_beta}), we see that the case with $\phi_0=3\pi/2$ can be related to that with $\phi_0=\pi/2$ by the transformation $\beta\rightarrow \beta^{-1}$. This means that we can directly utilize the four $\beta$-solutions in $\phi_0=\pi/2$ case, i.e., $|\beta_1|\leq|\beta_2|=|\beta_3|\leq|\beta_4|$, to obtain the solutions in the $\phi_0=3\pi/2$ case as $|\beta_4|^{-1}\leq|\beta_3|^{-1}=|\beta_2|^{-1}\leq|\beta_1|^{-1}$, without any change of the spectrum $E$. Therefore the topological transition points in the two cases should be identical. We have numerically confirmed that $\gamma_c$ has the same dependence on the parameters $\Omega_0,\Omega_r$ as in the case of $\phi_0=\pi/2$.}

\noindent

{\it \bf{Data Availability}}

\noindent
The datas that support the results of this study are available from the corresponding author upon reasonable request.\

{\it \bf{References}}

{\bf Acknowledgements}\\
 The work is supported by the National Key Research and Development Program of China (2018YFA0307600, 2017YFA0304100), the National Natural Science Foundation of China (No.12074419, 12134015, 11974331), and the Strategic Priority Research Program of Chinese Academy of Sciences (No. XDB33000000).\\
 {\bf Author contributions}\\
 The project was initiated by X.C. and supervised by X.C. and W. Y.. L.Z. performed the numerical calculations and L. H. provided assistance on double checking the data. All authors contributed to analyzing the results and writing the manuscript.\\
 {\bf Competing interests}\\
The authors declare no completing interests.

\clearpage

\begin{widetext}

In this Supplementary Information we provide  the Raman transition  diagram for a specific atomic candidate, the existence of non-Hermitian skin effect (NHSE) and band topology, and the lattice dynamics under realistic external confinements.
~\\

{\bf{Supplementary Note 1:Detailed Raman-transition scheme for $^{173}$Yb}}
~\\

We illustrate the implementation of our scheme, using level structures of $^{173}$Yb atoms as a concrete example.
In previous experiments with $^{173}$Yb, both the optical Raman lattice~\cite{Jo1} and 1D spin-orbit coupling (SOC) with dissipation (or the non-Hermitian SOC)~\cite{Jo2} have been successfully implemented.
The Raman transitions in these experiments are between the two hyperfine states in the ground-state $^{1}S_0$ manifold ($|\uparrow\rangle= |F=5/2, m_F=5/2\rangle$ and  $|\downarrow\rangle= |F=5/2, m_F=3/2\rangle$), and the excited states with $F'=(7/2, 5/2, 3/2)$ states the $^{3}P_1$ manifold. The spin-dependent loss is controlled by a near-resonant transition from $|\downarrow\rangle$ to the excited state $|F',m'_F\rangle=|7/2,1/2\rangle$. In view of these existing implementations, in Fig.~\ref{fig_setup_Yb} we show a specific Raman transition diagram based on the level structures of $^{173}$Yb, for realizing our scheme in Fig.~1(a) of the main text.

\begin{figure}[h]
\includegraphics[width=8cm]{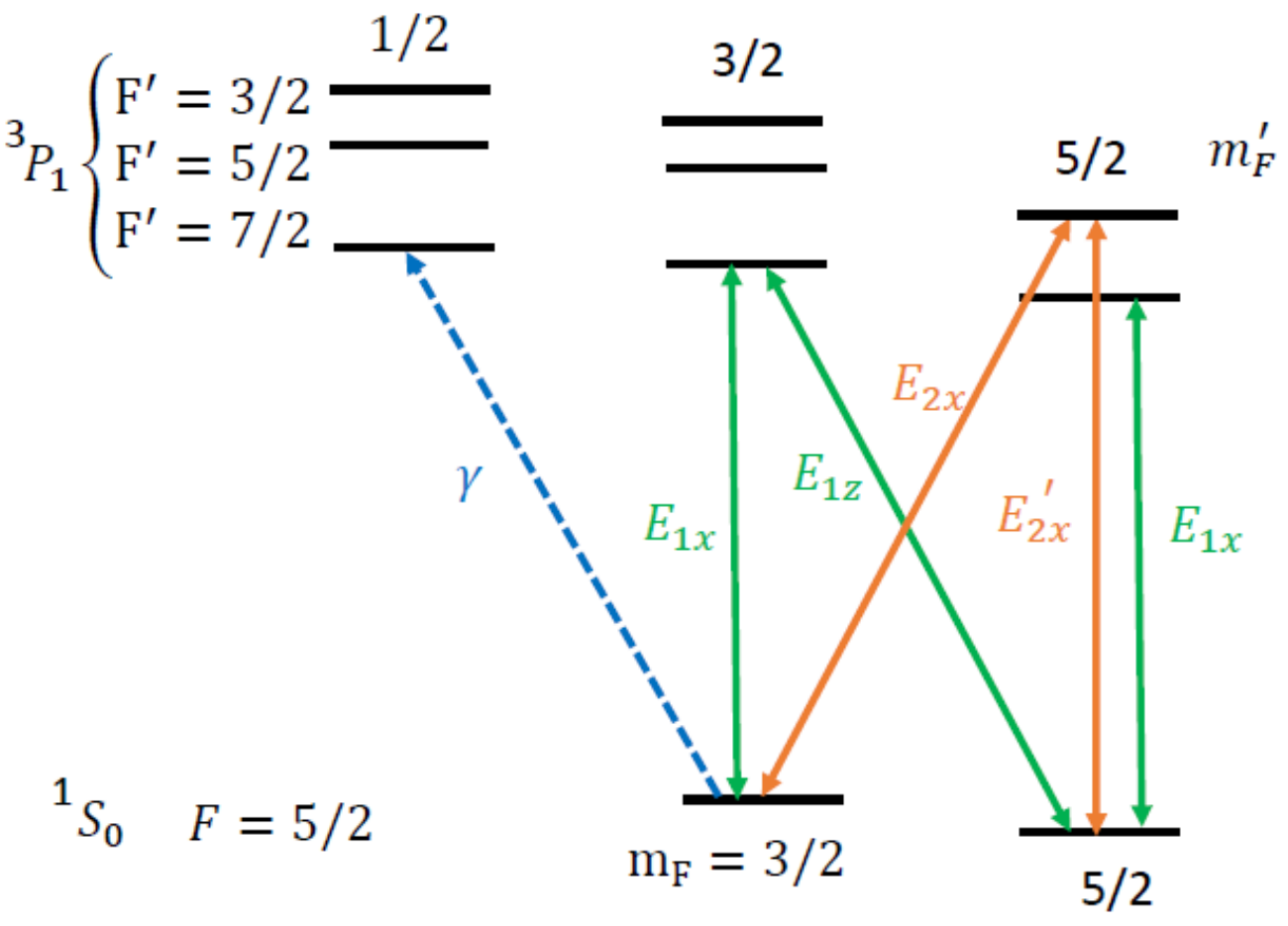}
\caption{(Color online) A proposed Raman-transition diagram for $^{173}$Yb. }
\label{fig_setup_Yb}
\end{figure}

Here the choice of two spin states $|\uparrow,\downarrow\rangle$, and the energy levels to create the optical Raman lattice, the $M_0$-SOC and the spin-dependent loss are all consistent with~\cite{Jo1,Jo2}. To ensure an independent control on the $M_r$-SOC (from the $M_0$-SOC), we propose to consider the Raman transition to a different excited state, say, $|F',m'_F\rangle=|5/2,5/2\rangle$, as shown in { Supplementary Figure 1}. A specific atomic transition can be achieved by tuning the laser frequency via the acousto-optical modulater (AOM) to exactly match the required atomic energy difference. 
~\\

{\bf{Supplementary Note 2:Non-Hermitian skin effect and band topology in the general $(\phi_r,\phi_0)$ phase plane}}
~\\

Here we discuss the presence of NHSE and band topology in case-III with general phase parameters $(\phi_r,\phi_0)$.

First, we discuss the band topology when the system is away from the points ($\phi_r=\pi,\phi_0=\pi/2,3\pi/2$). In Fig.~\ref{fig_spectrum}, we show the OBC spectrum by taking the same $(\Omega_0,\Omega_r,\gamma)$ as those in Fig.~3 of the main text, but with two different sets of phase parameters $(\phi_r,\phi_0)$. We see that the topological zero modes are now replaced by two finite-energy in-gap modes.

\begin{figure}[h]
\includegraphics[width=10cm]{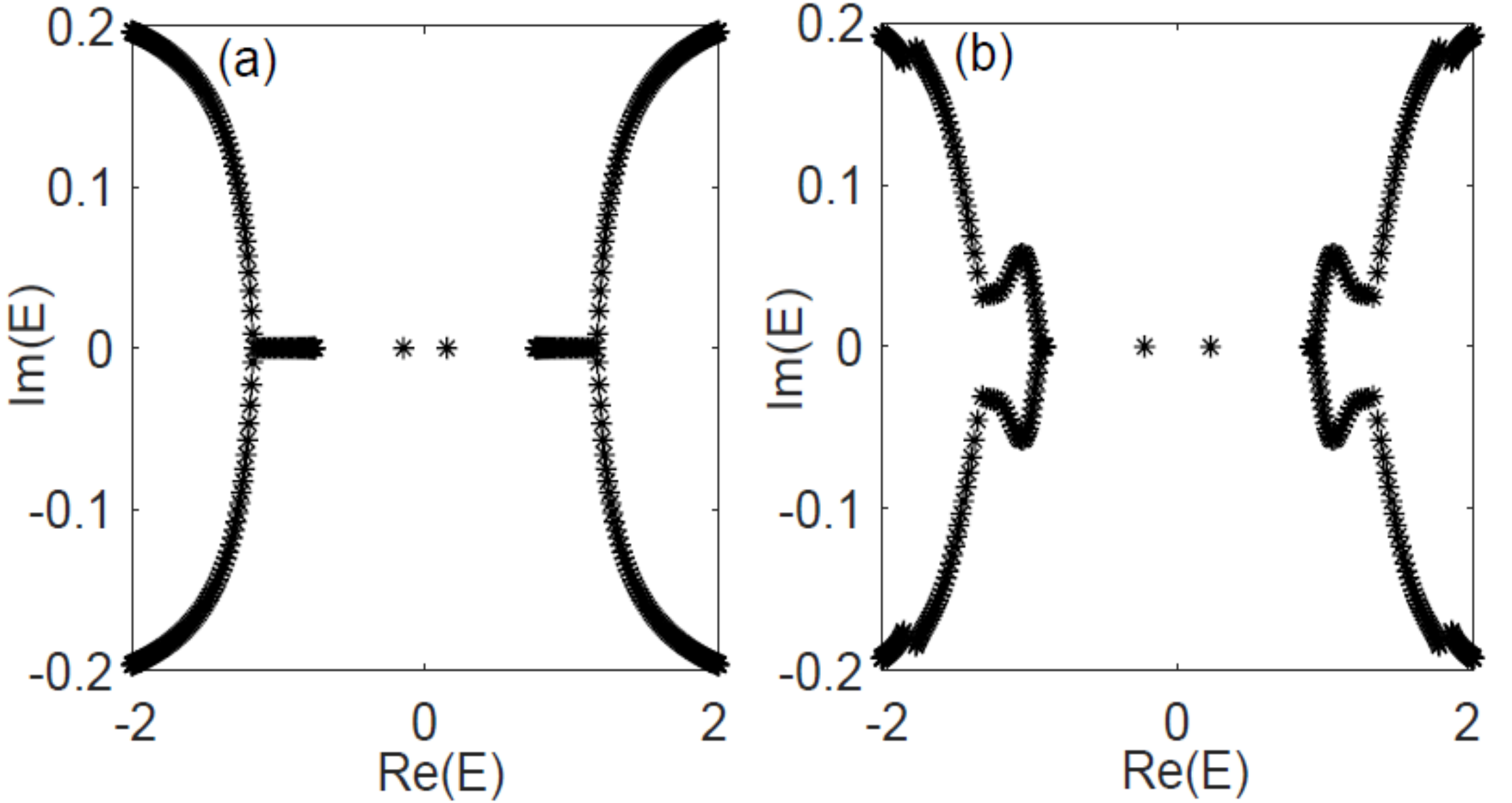}
\caption{ Energy spectrum under the open boundary condition (OBC) for (a) $(\phi_r,\phi_0)=(\pi,\pi/3)$ and (b) $(\phi_r,\phi_0)=(2\pi/3,\pi/2)$. In both cases we take $\Omega_0=0.5, \Omega_r=0.3, \gamma=0.2$, the same as in Fig.~3 of the main text. The energy unit is taken as hopping $t$.}
\label{fig_spectrum}
\end{figure}

To gain more insight to the vanishing of the topological zero modes under OBC, let us examine the Bloch Hamiltonian $H(k)$ for $\phi_r=\pi$ and $\phi_0\neq \pi/2, 3\pi/2$. In this case, $H(k)$ develops a $\sigma_y$ component, and the chiral symmetry is no longer present.
We have also verified from the tight-binding model that the chiral symmetry breaks down also for $\phi_r\neq\pi$, and thus the band topology is destroyed as long as $(\phi_r,\phi_0)$ deviates from the points $(\pi,\pi/2)$ and $(\pi,3\pi/2)$.

Finally, the appearance of the NHSE is found to be much more robust as compared to the band topology. We have explained in the main text that for the case $\phi_r=\pi$, the skin effect can hold for all $\phi_0$ except for $\phi_0=0,\pi$. For other $\phi_r\neq \pi$, we have numerically checked the energy spectrum under PBC and the eigen-modes under OBC, and conclude that the skin effect can expand to all the phase parameters except for $\phi_r=0$. In fact, for $\phi_r=0$, we can write down the tight-binding Hamiltonian in the $k$ space as
\begin{align}
H=&-\sum_k 2t\cos{k}(c_{k\uparrow}^\dagger c_{k\uparrow}-c_{k\downarrow}^\dagger c_{k\downarrow})+
i\gamma\sum_k(c_{k\uparrow}^\dagger c_{k\uparrow}-c_{k\downarrow}^\dagger c_{k\downarrow})\nonumber\\
&+\Omega_0\sum_k(-ie^{i\phi_0}2\sin{k}c_{k\uparrow}^\dagger c_{k\downarrow}+H.c.)+\Omega_r\sum_k(c_{k\uparrow}^\dagger c_{k-\pi,\downarrow}+H.c.).
\end{align}
We see that the last term couples $k$ to $k-\pi$, and in the basis $\psi=( c_{k-\pi,\uparrow},c_{k-\pi,\downarrow},c_{k,\uparrow},c_{k,\downarrow})^{T}$, we have $H=\sum_k \psi^{\dag} h(k) \psi$ with $h(k)$ a $4\times 4$ matrix. The diagonalization of $h(k)$ gives rise to the eigen-spectrum under PBC. In Fig.~\ref{fig_spectrum2}, we show two typical energy spectra in the complex energy plane for $\gamma<\Omega_r$ and $\gamma>\Omega_r$ with $\phi_0=\pi/2$. We see that the PBC spectrum in the complex plane behaves either as two disconnected lines or as two open arcs, but not closed loops. Other $\phi_0$ cases shows similar features. Therefore the NHSE does not show up if $\phi_r=0$.

\begin{figure}[h]
\includegraphics[width=10cm]{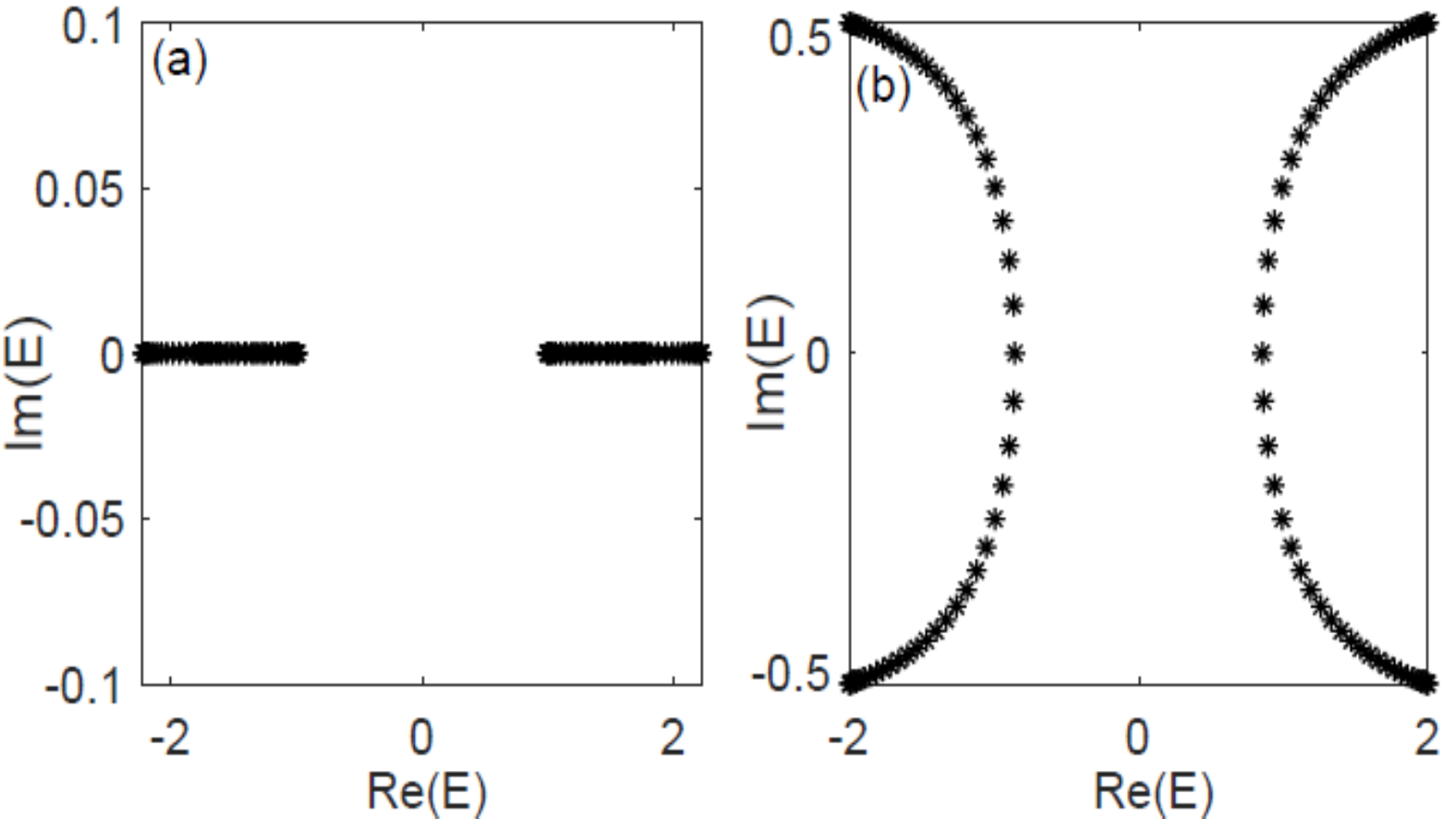}
\caption{ Energy spectrum under periodic boundary condition (PBC) for (a) $\Omega_0=0.5, \Omega_r=0.3, \gamma=0.2$ and (b)  $\Omega_0=0.5, \Omega_r=0.3, \gamma=0.6$. In both cases we take  $(\phi_r,\phi_0)=(0,\pi/2)$. The energy unit is taken as hopping $t$.}
\label{fig_spectrum2}
\end{figure}

\begin{figure}[h]
\includegraphics[width=8cm]{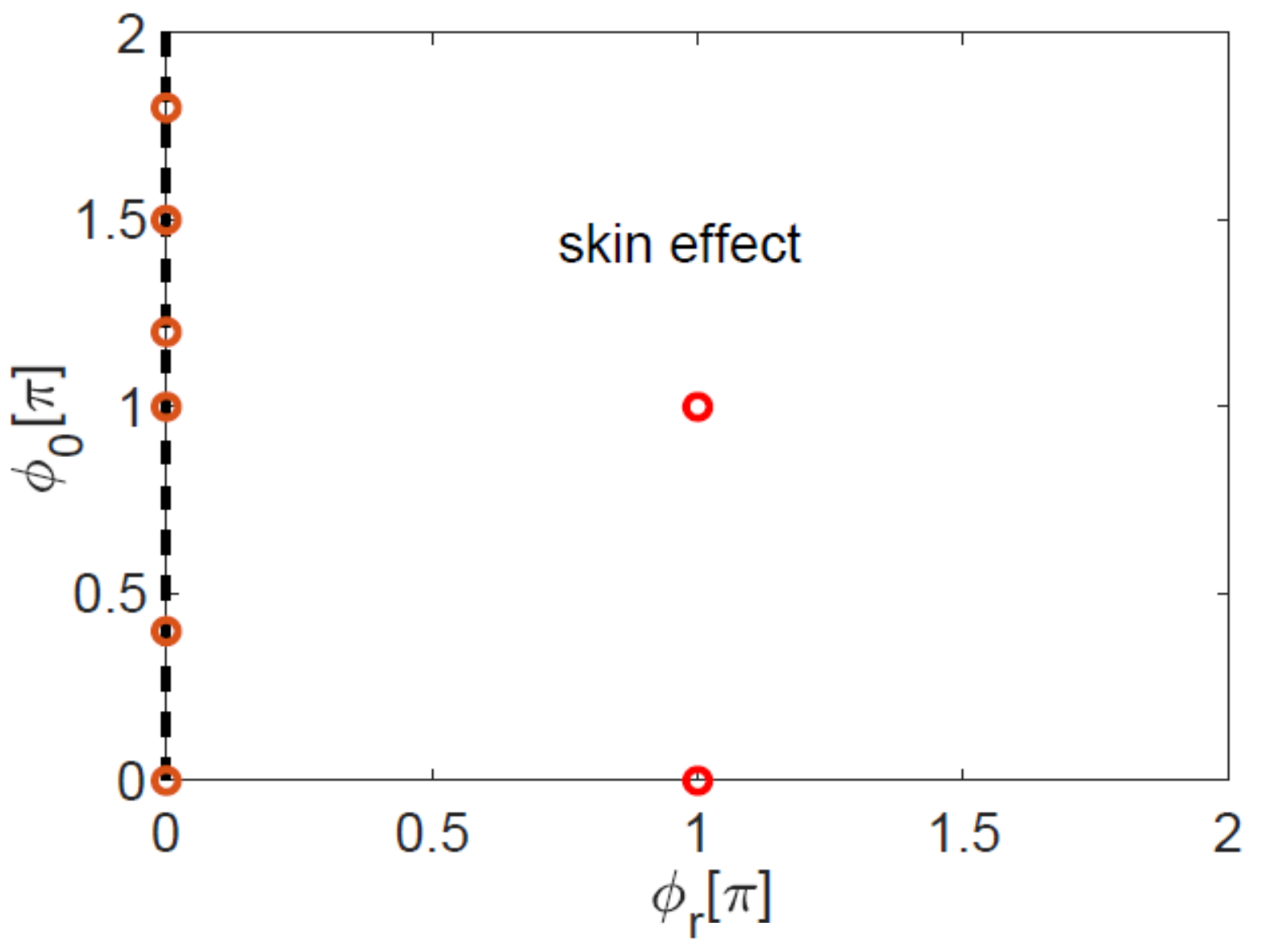}
\caption{(Color online) The region of non-Hermitian skin effect(NHSE) in the plane of $(\phi_r,\phi_0)$. The red circles at $(\pi,0)$ and $(\pi,\pi)$, as well as the dashed line at $\phi_r=0$, denote the parameter regimes for the absence of the non-Hermitian skin effect. The energy unit is taken as hopping $t$.}
\label{fig_phase}
\end{figure}

In Fig.~\ref{fig_phase} we summarize the existence regions for the NHSE for case-III in the phase plane  ($\phi_r,\phi_0$). It is shown that the skin effect is absent for two discrete points $(\pi,0)$ and $(\pi,\pi)$ and a line at $\phi_r=0$.

~\\

{\bf{Supplementary Note 3:Non-Hermitian skin effect under realistic confinements }}
~\\

Here we consider two types of realistic confinements in cold atoms, one is the box-trap potential with clear edges, and the other is the harmonic potential without a clear edge. In the first case, we consider the realistic situation when the edge is not sharp enough, i.e., there is an additional large but finite laser potential on each side of the lattice, see the illustration in Fig.~\ref{fig_box}(a). To examine the effect of edge imperfection, we have computed the edge-to-edge transport from $i=1$ to $L$ ($P_{LR}$) or from $i=L$ to $1$ ($P_{RL}$), when the laser potentials are applied at sites $i=0$ and $L+1$ with strength $V$. For simplicity, in our numerics the hopping rate between different sites and all other parameters are taken
to be the same as those without the box potential. The results are shown in Fig.~\ref{fig_box}((b)i,(b)ii) for $V=5t$.

\begin{figure}[h]
\includegraphics[width=13cm]{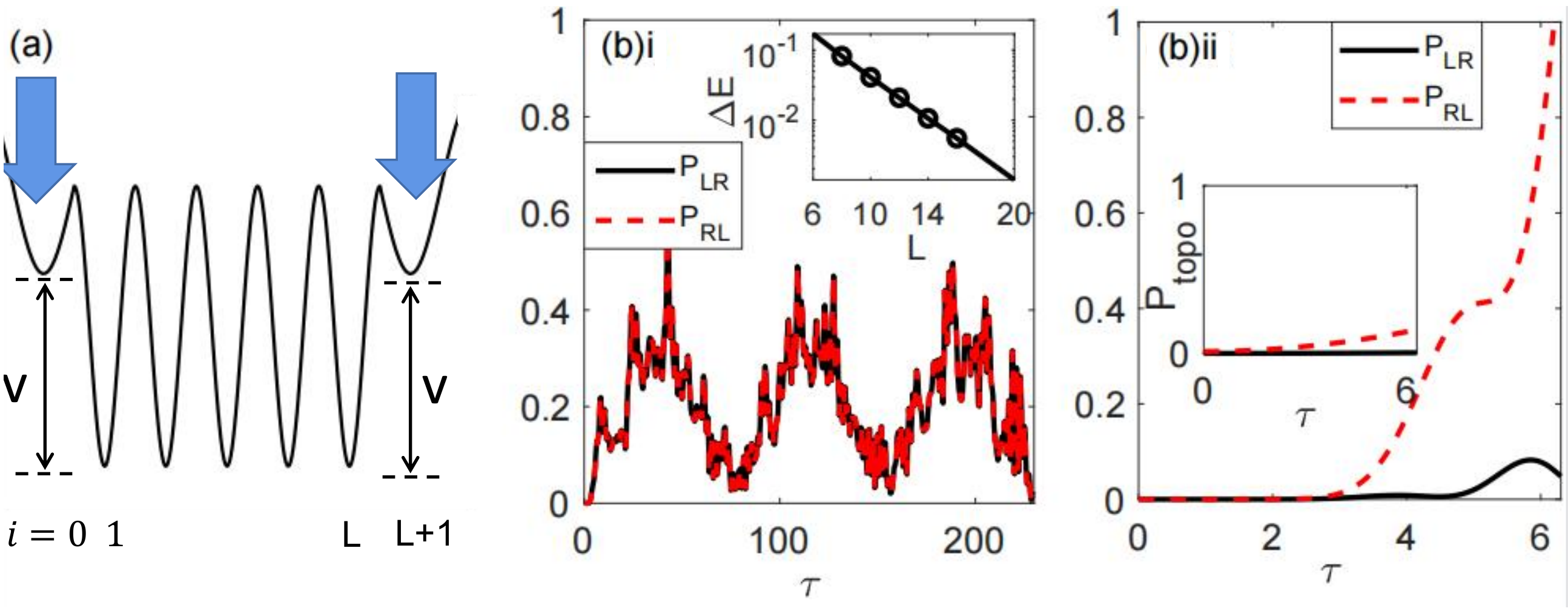}
\caption{(Color online) (a) Illustration of the lattice system under a box potential with imperfect edges. ((b)i, (b)ii) $P_{LR}$ and $P_{RL}$ between lattice sites $i=1$ and $L$ without ((b)i) and with ((b)ii) non-Hermitian skin effect(NHSE). Here we take the laser potential $V=5t$, and all other parameters are identical to Fig.~5 in the main text. The insets of ((b)i) and ((b)ii) are also the same physical quantities as in the insets of Fig.~5 in the main text. The energy unit is taken as hopping $t$, and the time unit is $\hbar/t$.}
\label{fig_box}
\end{figure}

For the first case, we see that the edge-to-edge transport shares qualitatively the same features as those in Fig.~5 of the main text. In the topological phase without NHSE (Fig.~\ref{fig_box}((b)i)), $P_{LR}$ and $P_{RL}$ are identical to each other, and the oscillation frequency decays exponentially with size $L$. In comparison, the dynamics with NHSE shows obvious directional preference (Fig.~\ref{fig_box} ((b)ii)), as the localized bulk states contribute significantly.

We now turn to the second case and discuss the manifestation of NHSE in a lattice system under a harmonic trap, $V(x)=V_0x^2$, where the edge is not well-defined.  In our simulation, we take the trap center at the lattice site $i_0$ and thus there is a local on-site potential $\mu_i=V(i-i_0)^2$ to each site index $i$, with $V=V_0a^2$ ($a$ is the lattice spacing). We then consider the expansion dynamics of a particle from the trap center $i=i_0$, see illustration in Fig.~\ref{fig_har}(a). To eliminate the difference caused by spin, we consider the initial state as an equal population of $|\uparrow\rangle$ and $|\downarrow\rangle$ at site $i_0$, and then examine the probability of finding a particle at site $n$ after an evolution time of $\tau$, 
defined as
\begin{equation}
\rho_n(\tau) \equiv \sum_{\sigma}|\psi_{n,\sigma}(\tau)|^2. 
\end{equation}

\begin{figure}[h]
\includegraphics[width=13cm]{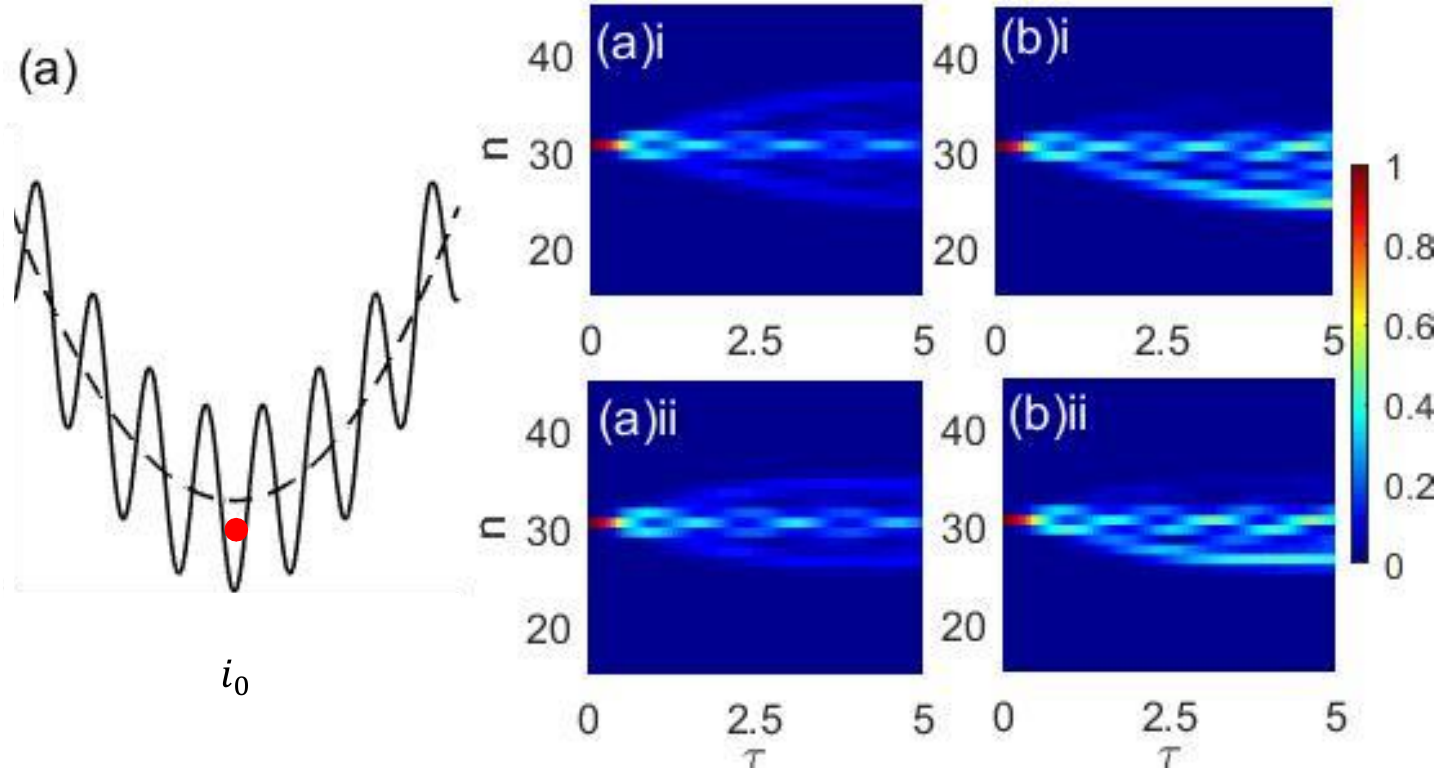}
\caption{(Color online) (a) Illustration of the lattice system with an additional harmonic potential. ((a)i,(a)ii,(b)i,(b)ii) Contour plot of $\rho_n(\tau)$ in the ($n,\tau$) plane for the bulk dynamics in a harmonic trap. Here we take $\Omega_r=1,\Omega_0=0.5$, and other parameters are $(\gamma, V)=(0,0.1)$((a)i),  $(0.2,0.1)$((b)i), $(0,0.2)$((a)ii), $(0.2,0.2)$((b)ii). The lattice size is $L=60$, and the particle is initialized at $i_0=L/2$ (the trap center). The energy unit is taken as hopping $t$, and the time unit is $\hbar/t$.
}
\label{fig_har}
\end{figure}

In Fig.~\ref{fig_har}((a)i,(a)ii,(b)i,(b)ii), we show the contour plot of $\rho_n(\tau)$ in the ($n,\tau$) plane for various cases of $\gamma$ and $V$ parameters.  One can see that in the absence of NHSE ($\gamma=0$ in Fig.~\ref{fig_har}((a)i,(a)ii)), the particle wave function
expands from the trap center with equal probabilities to the right- and left-hand sides. In comparison, when the NHSE is present under a finite $\gamma$ in Fig.~\ref{fig_har}((b)i,(b)ii), the expansion shows strong directional preference towards the left-hand side (smaller index) of the lattice. The probability peak apparently moves to the left-hand side of the trap center at short times, which then saturates at a certain distance to the trap center at longer times.
By comparing Fig.~\ref{fig_har}((a)i,(a)ii) and ((b)i,(b)ii) with different harmonic confinements, we can see that a stronger $V$ will suppress the directional flow, i.e., the off-center displacement of the particle becomes smaller as $V$ increases, whereas the off-center accumulation of population becomes more appreciable.
We have also checked that in the absence of the harmonic trap $V=0$, the expansion peak moves to left-hand side continuously (before getting close to any boundaries). These results show that even there is no sharp edge or boundary, the NHSE can also manifest itself in the bulk dynamics.

{{\bf{Supplementary References}}}

\end{widetext}

\end{document}